\documentclass[sigconf]{acmart}

\AtBeginDocument{%
  \providecommand\BibTeX{{%
    \normalfont B\kern-0.5em{\scshape i\kern-0.25em b}\kern-0.8em\TeX}}}
\usepackage{diagbox}
\usepackage{booktabs}
\usepackage{adjustbox}
\usepackage{color}
\usepackage{multirow}
\usepackage{graphicx}
\usepackage{ulem}
\graphicspath{ {./pic/} }
\setcopyright{acmcopyright}
\copyrightyear{2025}
\acmYear{2025}
\acmDOI{XXXXXXX.XXXXXXX}

\acmConference[XXX]{xxx}{xxx}

\begin{document}

\title{ULMRec: User-centric Large Language Model for  Sequential Recommendation}

\author{Minglai Shao}
\affiliation{%
  \institution{School of New Media and Communication, Tianjin University, Tianjin, China}
  \city{}
  \state{}
  \country{}
}
\email{shml@tju.edu.cn}

\author{Hua Huang}
\affiliation{%
  \institution{School of New Media and Communication, Tianjin University, Tianjin, China}
  \city{}
  \state{}
  \country{}
}
\email{huanghua18@tju.edu.cn}

\author{Qiyao Peng}
\authornote{Corresponding author}
\affiliation{%
  \institution{School of New Media and Communication, Tianjin University, Tianjin, China}
  \city{}
  \state{}
  \country{}
}
\email{qypeng@tju.edu.cn}

\author{Hongtao Liu}
\affiliation{%
  \institution{Du Xiaoman Financial}
  \city{Beijing, China}
  \state{}
  \country{}
}
\email{liuhongtao01@duxiaoman.com}

\begin{abstract}
Recent advances in Large Language Models (LLMs) have demonstrated promising performance in sequential recommendation tasks, leveraging their superior language understanding capabilities. 
However, existing LLM-based recommendation approaches predominantly focus on modeling item-level co-occurrence patterns while failing to adequately capture user-level personalized preferences. 
This is problematic since even users who display similar behavioral patterns (e.g., clicking or purchasing similar items) may have fundamentally different underlying interests.
To alleviate this problem, in this paper, we propose ULMRec, a framework that effectively integrates user personalized preferences into LLMs for sequential recommendation.
Considering there has the semantic gap between item IDs and LLMs, we replace item IDs with their corresponding titles in user historical behaviors, enabling the model to capture the item semantics.
For integrating the user personalized preference, we design two key components: (1) user indexing: a personalized user indexing mechanism that leverages vector quantization on user reviews and user IDs to generate meaningful and unique user representations, and (2) alignment tuning: an alignment-based tuning stage that employs comprehensive preference alignment tasks to enhance the model's capability in capturing personalized information.
Through this design, ULMRec achieves deep integration of language semantics with user personalized preferences, facilitating effective adaptation to recommendation.
Extensive experiments on two public datasets demonstrate that ULMRec significantly outperforms existing methods, validating the effectiveness of our approach.
\end{abstract}


\maketitle

\section{Introduction}

Recently, recommender systems have emerged as a cornerstone technology across digital platforms, playing a pivotal role in filtering and suggesting relevant content from vast information spaces to meet users' individual preferences~\cite{schafer2007collaborative}. The dynamic nature of user interests and behavioral patterns has highlighted the limitations of conventional recommendation methods, prompting increased attention on sequential recommendation approaches~\cite{sasrec,bert4rec}, which could effectively capture user temporal interests.

Most existing sequential recommendation methods adopt various of neural networks to model user historical behavior to capture the item co-occurrence patterns, such as RNN~\cite{narm}, CNN~\cite{tang2018personalized,yuan2019simple}, Transformer~\cite{sasrec}, etc.
SASRec~\cite{sasrec} uses a unidirectional Transformer to model item sequences and predict the next item, combining features of Markov Chains and RNNs with self-attention.
Recently, Large Language Models (LLMs)~\cite{touvron2023llama} have become increasingly prevalent in recommender systems~\cite{llamarec,lcrec} due to their ability to understand and generate human-like natural language.
For example, 
PALR~\cite{palr} uses LLMs to generate the user profile and pre-filter the candidate item pool, then fine-tunes the LLMs to recommend.
LlamaRec~\cite{llamarec} employs the LLMs to re-rank the candidate items retrieved by the small-scale sequential recommendation methods.
LC-Rec~\cite{lcrec} indexes the items with discrete IDs and injects them into LLMs via various of alignment tasks.
Despite the remarkable achievements of LLMs in recommendation tasks, we identify a critical limitation: these models predominantly focus on modeling item-level co-occurrence patterns while failing to capture user-level personalization needs.
A key challenge is that users may interact with identical items yet have entirely different motivations and interests, a distinction that current LLM-based approaches do not adequately address.

To bridge this gap, we propose developing an effective framework that incorporates user personalized preferences into LLMs. 
A straightforward approach would be to integrate user IDs, which serve as unique user identifiers, into LLMs by prefixing them to users' historical behaviors during the fine-tuning process. 
However, this naive integration faces two significant challenges:

\textit{(1) Semantic gap.} A significant disconnect exists between the language semantics modeled by LLMs and the preference information embedded in user IDs within recommender systems.
This disconnect occurs because LLMs are trained on natural language and may not recognize the special significance of user IDs in the context of recommendation.
When LLMs tokenize user IDs, they inadvertently fragment and potentially destroy the inherent personalized preference information encoded within these identifiers.

\textit{(2) Limited task.} Fine-tuning LLMs solely on the next-item prediction task may constrain the model to learn merely superficial item co-occurrence patterns in users' historical sequences, rather than developing a comprehensive understanding of users' personalized preferences. 
This narrow focus would not capture the fine-grained user personalized preferences, which extend beyond simple sequential patterns.

Hence, we tackle this integration problem in two main aspects:

(1) User Indexing: we develop a personalized allocation mechanism that generates meaningful and unique user indices. These indices must satisfy two critical criteria: effectively capturing user preferences; ensuring distinct representations without allocation conflicts.

(2) Alignment Tuning: we design various of alignment strategies beyond simple next-item prediction task to achieve: deep integration of language semantics with user preferences within LLMs; comprehensive modeling of user preferences beyond simple behavioral patterns.

In this paper, we propose ULMRec, an LLM-based model to incorporate the language semantics and user personalized preference for improving recommendation.
Firstly, considering the LLMs could not capture the item ID semantic features, we directly employ the item title to replace that in the user historical behaviors.
For integrating the user personalized preference, our model has two main phases: personalized user indexing stage and alignment-based user index understanding stage. 
For the first stage, since the user's historical reviews contain rich information about the user's interests, we extract the user index from the corresponding reviews. 
In detail, we propose to employ the vector quantization~\cite{lee2022autoregressive} towards the user historical reviews to generate the personalized user index.
To mitigate potential index collisions, we incorporate original user IDs into the quantization process, which could ensure each user index is both semantically meaningful and unique.
While our generated indices contain valuable personalized information and avoid semantic conflicts with the LLM's token IDs, the challenge lies in enabling the LLM to interpret these indices accurately. 
To address this, we design a comprehensive set of preference alignment tasks, beyond traditional sequential recommendation task. 
These tasks are crafted to enhance the model's capability in capturing the personalized information embedded within the user indices.
In this way, the user-level personalized information can be injected into the LLMs for improving recommendation.

In all, the contributions of our paper can be summarized as below:
\begin{itemize}
    \item We propose ULMRec, which could bridge the semantic gap between LLMs and personalized recommendation by integrating user preferences into LLMs. It is the first attempt to address the fundamental challenge of incorporating user-level personalization into LLMs.
    \item We develop an innovative personalized user indexing mechanism for generating meaningful and unique user indexes. Besides, combined with our carefully designed alignment task, the LLMs could understand the user personalized preferences deeply rather than merely learning superficial item co-occurrence patterns.
    \item Extensive experiments on two public datasets demonstrate the effectiveness of our approach in integrating user personalized preferences into LLMs.
\end{itemize}

\section{Related Work}

\subsection{Large Language Model}

Large Language Models (LLMs)~\cite{liu2024bucket} have revolutionized natural language processing in recent years. 
The development of LLMs can be traced through several key milestones. Initially, models like BERT~\cite{bert} introduced the concept of pre-training on vast amounts of text data, enabling better understanding of language context.
A significant leap came with GPT-3~\cite{brown2020language}, which demonstrated remarkable few-shot learning capabilities across various tasks. 
This was followed by instruction-tuned models like InstructGPT~\cite{ouyang2022training}, which improved alignment with human intent. More recent models such as Llama~\cite{touvron2023llama} have pushed the boundaries of model size and efficiency, achieving state-of-the-art performance on numerous benchmarks.
The applications of LLMs have expanded rapidly, ranging from text generation and summarization to more complex tasks like reasoning and code generation.

Despite their remarkable capabilities, LLMs face significant challenges when applied to personalized recommender systems. 
The reason is that these models excel at general language understanding and generation tasks, but they struggle to capture fine-grained user preferences crucial for effective recommendations.

\subsection{Sequential Recommendation}
Sequential recommendation aims to suggest items of interest to users by modeling the sequential dependencies over the user-item interactions. 
Early works in this area mainly employ the Markov Chain techniques~\cite{rendle2010factorizing,he2016fusing,wang2015learning}. 
With the development of deep learning, various kinds of neural network techniques are proposed to encode interaction sequences and capture the item co-occurrence patterns, such as RNN~\cite{narm}, MLP~\cite{zhou2022filter}, CNN~\cite{tang2018personalized,yuan2019simple}, GNN~\cite{wu2019session,chang2021sequential} and Transformer~\cite{sasrec,bert4rec,peng2024pept}.

Recently, Pre-trained Language Models (PLMs) like BERT~\cite{DBLP:conf/naacl/DevlinCLT19}, Llama~\cite{touvron2023llama} have demonstrated remarkable text comprehension and reasoning capabilities, prompting researchers to explore their integration into various fields~\cite{liu2022expertbert,peng2022expertplm}, such as recommender systems~\cite{tallrec,cui2022m6,palr,p5}.
A common approach involves representing user behaviors as text sequences and designing prompts to instruct LLMs in performing recommendation tasks. 
For instance, TallRec~\cite{tallrec} structured recommendation data as instructions, incorporating users' historical interactions and target items.
Recognizing the semantic gap between LLMs and recommender systems, some researchers have proposed item indexing methods and alignment tasks.
For example, LC-Rec~\cite{lcrec} employed a tree-structured vector quantization method to index items and used carefully designed alignment tasks to prompt the LLMs to understand the indexes.
CoLLM~\cite{collm} captured and mapped the collaborative information through external traditional models, forming collaborative embeddings used by LLMs.

Despite these advancements, current methods primarily focus on capturing item co-occurrence patterns in user-item interactions, often overlooking user-specific preference information. 
Our work aims to bridge this gap by integrating LLMs with traditional recommender systems from a user modeling perspective, thereby not only leveraging item co-occurrence patterns but also deeply understanding and incorporating personalized user preferences.

\section{Problem Statement}

Sequential recommendation aims to predict users' possible interactions based on their historical interactions. 
Formally, we define the problem as follows:
Let $\mathcal{U} = [{u_1, u_2, ..., u_{|\mathcal{M}|}}]$ denotes the set of users and $\mathcal{I} = [{i_1, i_2, ..., i_{|\mathcal{N}|}}]$ denotes the set of items, where $\mathcal{M}$ and $\mathcal{N}$ represent the number of users and items respectively.
For each user $u \in \mathcal{U}$, their interaction history is represented as a sequence $S_u = [i_1, i_2, ..., i_t]$, where $i_k \in \mathcal{I}$ and $t$ is the sequence length. 
Each interaction is associated with a timestamp, and the sequence is ordered chronologically.
Given a user $u$ and their interaction history $S_u$, the goal of sequential recommendation is to predict the next item $i_{t+1}$ that the user is likely to interact with.

\section{Proposed Method}
\begin{figure*}
    \centering
    \includegraphics[width=0.95\linewidth]{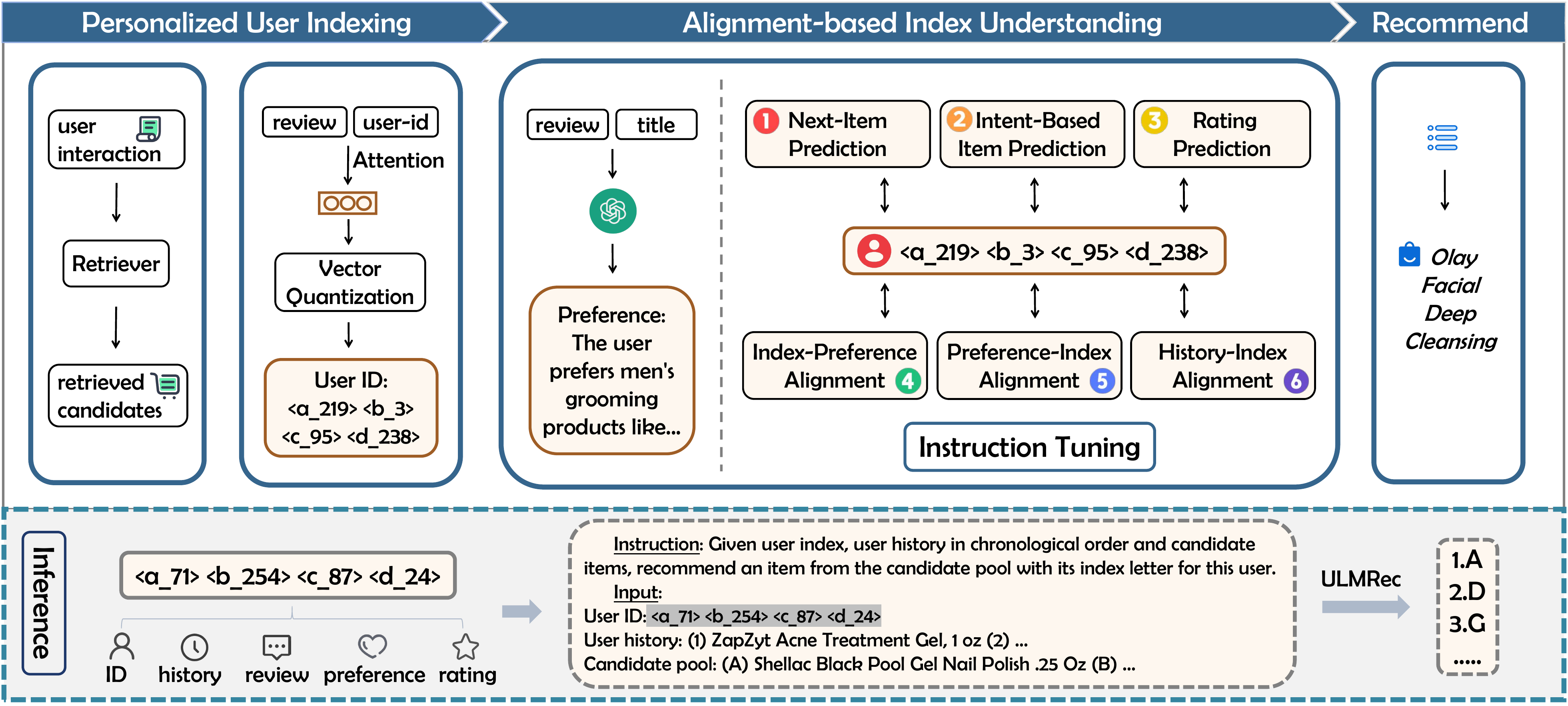}
    \caption{The framework of our ULMRec. We incorporate user-specific preferences into the LLMs through personalized user indexing and alignment-based index understanding.}
    \label{model}
\end{figure*}

In this section, we will introduce ULMRec in detail. 
As discussed above, there exists a large gap between
the language semantics modeled by LLMs and user personalized semantics implied by recommender systems, which limits the capacities of LLMs in recommendation. 
To effectively bridge this gap, as illustrated in Figure~\ref{model}, our framework consists of the following two stages:

\textbf{Personalized User Indexing Stage.} We integrate the user reviews and personalized information to generate the unique hierarchical semantic IDs for each user via vector quantization.
In this way, the learned user indices can capture similarities among different users, and provide a unique indexing representation for each specific user.

\textbf{Alignment-based Index Understanding Stage.} In addition to the traditional next-item prediction task, we instruction tune the LLM based on a comprehensive set of preference alignment task to establish connections between the LLMs and users' personalized information.

\subsection{Personalized User Indexing}
The key to introducing the user personalized preference into the LLMs for improving recommendation is how to represent the user with index IDs and integrate these IDs into LLMs.
Firstly, directly using original user IDs poses a risk of semantic conflicts within the LLM framework, as these IDs may inadvertently interact with or be misinterpreted by the model's existing token semantics.
Secondly, although constructing user profiles through LLM-generated descriptions offers an alternative approach, it often fails to capture the nuanced and individualistic aspects of user preferences. This method tends to produce generalized profiles that lack sufficient differentiation, potentially resulting in homogeneous representations that fail to reflect the diverse spectrum of user interests and behaviors.
In this section, we propose an approach that leverages vector quantization techniques~\cite{rqvae} to generate user indices. This method enables the encapsulation of personalized preference information in a format compatible with LLM architectures.

To generate semantic user indices that encapsulate rich preference information, we leverage user-generated reviews, which are often rich with insights into individual preferences and behaviors.
We initially employ BERT~\cite{bert}, a powerful pre-trained language model, to encode the textual content of user reviews. For each user $u \in \mathcal{U}$, their reviews are represented as a sequence $W_u = [w_1, w_2, ..., w_n]$, where $n$ denotes the length of historical reviews. The BERT model transforms these reviews into high-dimensional embeddings: $\mathbf{E}_u = BERT(W_u) = [\mathbf{e}_{w_1}, \mathbf{e}_{w_2}, ..., \mathbf{e}_{w_n}]$.
Although reviews already contain rich user preference information, we obtain the embedding $\mathbf{e}_{o_u}$ of user's original ID $o_u$ in the same way as reviews and integrate it using an attention mechanism to avoid collisions about the index. 
The attention process can be formulated as:
\begin{equation}
    \alpha_i = softmax(\mathbf{e}_{w_i}^T \mathbf{A}_i \mathbf{e}_{o_u}),
    \mathbf{x}_u = \sum_{i=1}^{n} \alpha_i \mathbf{e}_{w_i} ,
\end{equation}
where $\mathbf{A}_i$ is a learnable weight matrix, $\alpha_i$ is the attention weight for the $i$-th review, and $\mathbf{x}_u$ is the aggregated review representation.

Then, using it as input, we train a Residual-Quantized Variational AutoEncoder (RQ-VAE)~\cite{rqvae} to hash the user information into discrete personalized semantic IDs in a unified space. 
Specifically, RQ-VAE encodes the input user embedding $\mathbf{x}$ to obtain a latent representation $\mathbf{z}$. 
At the initial level ($l$ = 0), residual is defined as $\mathbf{z}$. And in each level $l$, we have a codebook $C_{}^{l}=\{\mathbf{v}_{k}^{l}\}_{k=1}^{K}$, where $\mathbf{v}_{k}^{l}$ is the codebook vector and $K$ is the codebook size. 
At the 0-th level, we map the latent representation $\mathbf{r}_{0}=\mathbf{z}$ to the closest vector $\mathbf{v}_{k}^{0}$ in codebook $C_{}^{0}$, where the index of it is the 0-th codeword $d_{0}$. Then, at the next level, the residual vector is computed as: $\mathbf{r}_{1}=\mathbf{r}_{0}-\mathbf{v}_{d_{0}}^{0}$. Similar to before, $d_{1}$ is obtained through finding the closest embedding in the codebook $C_{}^{1}$ which is nearest to $\mathbf{r}_{1}$. This process will be performed $p$ times iteratively to get $p$ codewords as the hierarchical semantic IDs.
Finally, the sum of each quantized codebook vector is an approximation of the original input vector. 
Below describes this process, where $d_{i}$ represents the index of the closest embedding, namely the $i$-th codeword of user indices.

\begin{equation}
d_{i}=\underset{k}{\arg \min }\left\|\mathbf{r}_{i}-\mathbf{v}_{k}^{i}\right\|_{2}^{2} ,
\end{equation}
\begin{equation}
\mathbf{r}_{i+1}=\mathbf{r}_{i}-\mathbf{v}_{d_{i}}^{i} .
\end{equation}

After generating the semantic IDs, the quantized representation of $\mathbf{z}$, computed as: $\mathbf{\hat{z}}=\sum_{i=0}^{p-1}{\mathbf{v}_{d_i}^{i}}$, is used as the decoder input to re-construct the input user embedding $\mathbf{x}$. 
The training loss contains reconstruction loss and RQ loss, which are defined as follows:

\begin{equation}
\mathcal{L}_{\mathrm{RECON}} =\|\mathbf{x}-\mathbf{\hat{x}}\|_{2}^{2} ,
\end{equation}
\begin{equation}
\mathcal{L}_{\mathrm{RQ}} =\sum_{i=0}^{p-1}\left\|\operatorname{sg}\left[\mathbf{r}_{i}\right]-\mathbf{v}_{d_{i}}^{i}\right\|_{2}^{2}+\beta\left\|\mathbf{r}_{i}-\operatorname{sg}\left[\mathbf{v}_{d_{i}}^{i}\right]\right\|_{2}^{2} ,
\end{equation}
\begin{equation}
\mathcal{L}_{\mathrm{RQ}-\mathrm{VAE}} =\mathcal{L}_{\mathrm{RECON}}+\mathcal{L}_{\mathrm{RQ}} .
\end{equation}

\subsection{Alignment-based Index Understanding}
Although user semantic IDs are constructed, the LLMs often struggle to fully grasp their intrinsic meanings. 
To better integrate the user semantic IDs into the LLMs, we design a series of customized tasks aimed at aligning these indices with personalized user semantics. 
This approach encourages LLMs to comprehensively learn about the underlying personal preferences from provided context.
To mitigate the issue of catastrophic forgetting in LLMs, we employ a strategy of shuffling the task data and randomly selecting samples from each task according to specified proportions. 
This approach ensures a balanced and diverse learning for the models.

\subsubsection{Next-Item Prediction}
\label{sec:seq}
\begin{figure}
    \centering
    \includegraphics[width=0.82\linewidth]{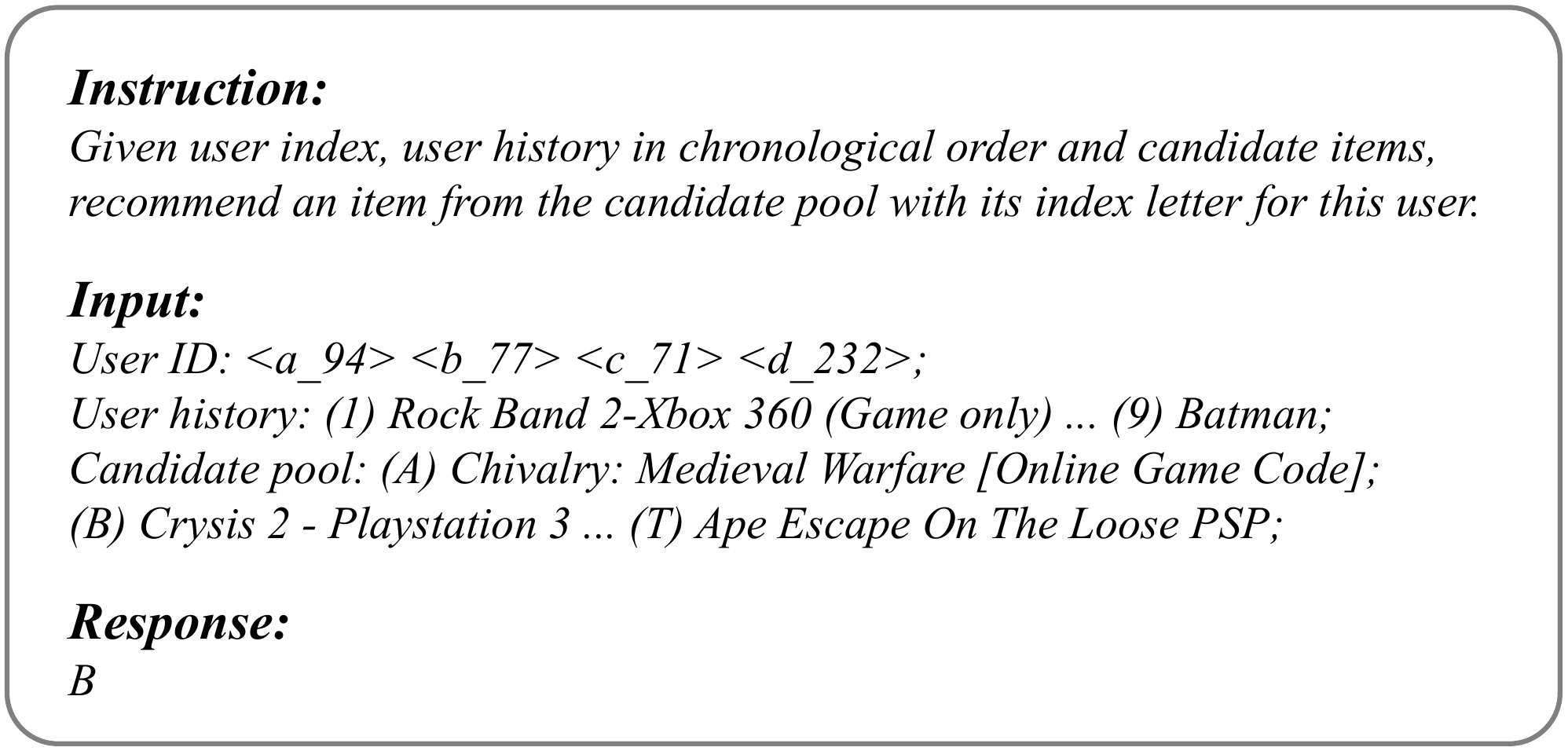}
    \caption{Next-item prediction task.}
    \label{nextitem}
\end{figure}

Our primary instruction tuning task focuses on next-item prediction. 
We construct personalized recommendation prompts by combining three key elements: the user index, a chronological sequence of historical interactions, and a set of previously retrieved candidate items. 
LLMs are enforced to predict the next item that a user is most likely to interact with from the candidate item pool.
Figure~\ref{nextitem} is an example prompt.

However, relying solely on this task could not sufficiently integrate the user personalized preferences. 
Additional deep personalized semantic alignment tasks are necessary. 

\subsubsection{Index-Preference Alignment}
\label{sec:pref}
\begin{figure}
    \centering    \includegraphics[width=0.82\linewidth]{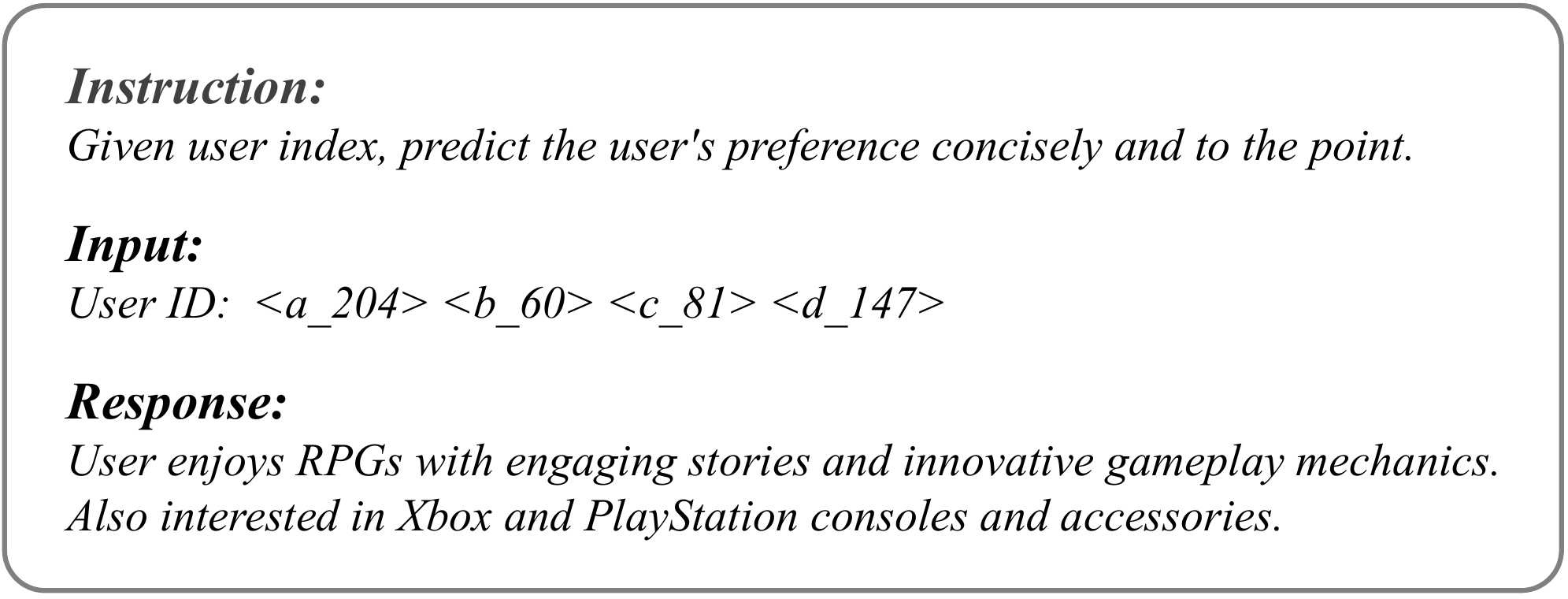}
    \caption{Index-preference alignment task.}
    \label{indexpref}
\end{figure}

Explicitly aligning user indices with user preference information is crucial for our model. 
To achieve this, we employ two processes:
(1) Preference extraction: We utilize GPT-3.5-Turbo~\cite{brown2020language} to extract user preferences from reviews, since reviews often reflect users' attitudes and tastes comprehensively;
(2) Index-preference alignment: To enhance LLMs' understanding and inference capabilities regarding user indices, we instruct them to reconstruct user preferences based solely on the user index.
In this way, the LLM could effectively interpret and utilize these indices in subsequent tasks, grounding them in actual user preferences and behaviors.
Figure~\ref{indexpref} is a sample instance.

\subsubsection{Preference-Index Alignment}
\label{sec:turnpref}
\begin{figure}
    \centering
    \includegraphics[width=0.83\linewidth]{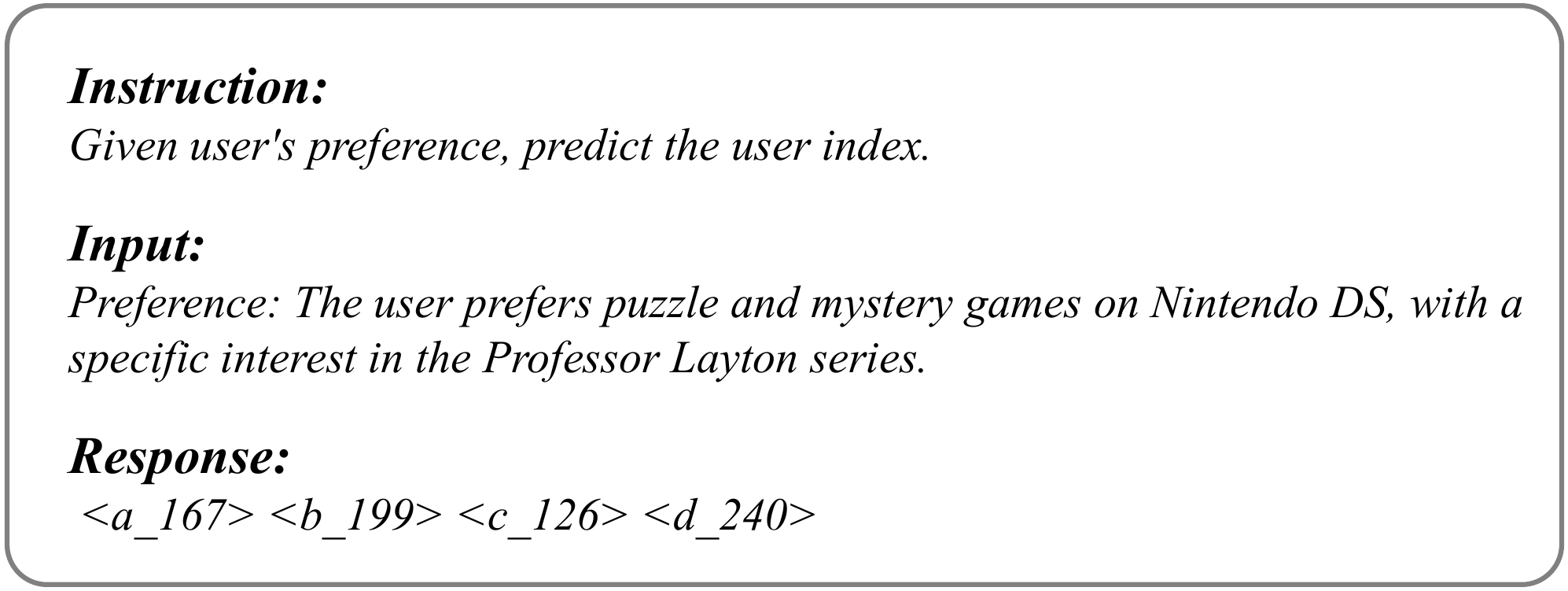}
    \caption{Preference-index alignment task.}
    \label{prefindex}
\end{figure}

In order to further enhance the LLMs' ability to understand and include user preferences, we reverse the input and output of the index-preference alignment task.
This reversed task serves as a counterpart to the earlier index-to-preference mapping, creating a bidirectional understanding of the relationship between user indices and preferences.
In this way,  the model could better associate user preferences with their respective indices in a bidirectional learning process.
Figure~\ref{prefindex} shows an example prompt.

\subsubsection{History-Index Alignment}
\label{sec:history}
\begin{figure}
    \centering
    \includegraphics[width=0.82\linewidth]{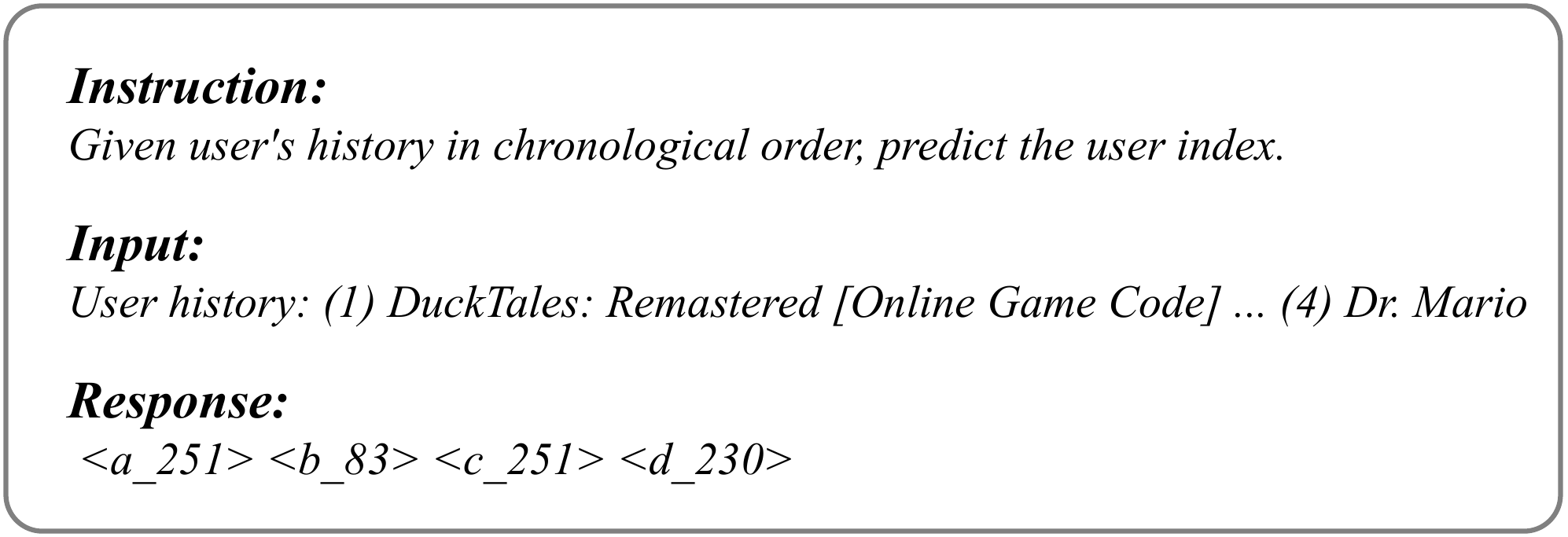}
    \caption{History-index alignment task.}
    \label{historyindex}
\end{figure}

Users with similar interaction histories may have distinct preferences, for example, one user might prioritize price while another focuses on quality. To address this nuance, we design a task that aligns historical behavior with unique user indices. This alignment aims to couple each user's interaction history with their distinctive index, enabling the LLMs to differentiate between users with similar behaviors but divergent preferences.
Figure~\ref{historyindex} is an example of this task.

\subsubsection{Rating Prediction}
\label{sec:rating}
\begin{figure}
    \centering
    \includegraphics[width=0.82\linewidth]{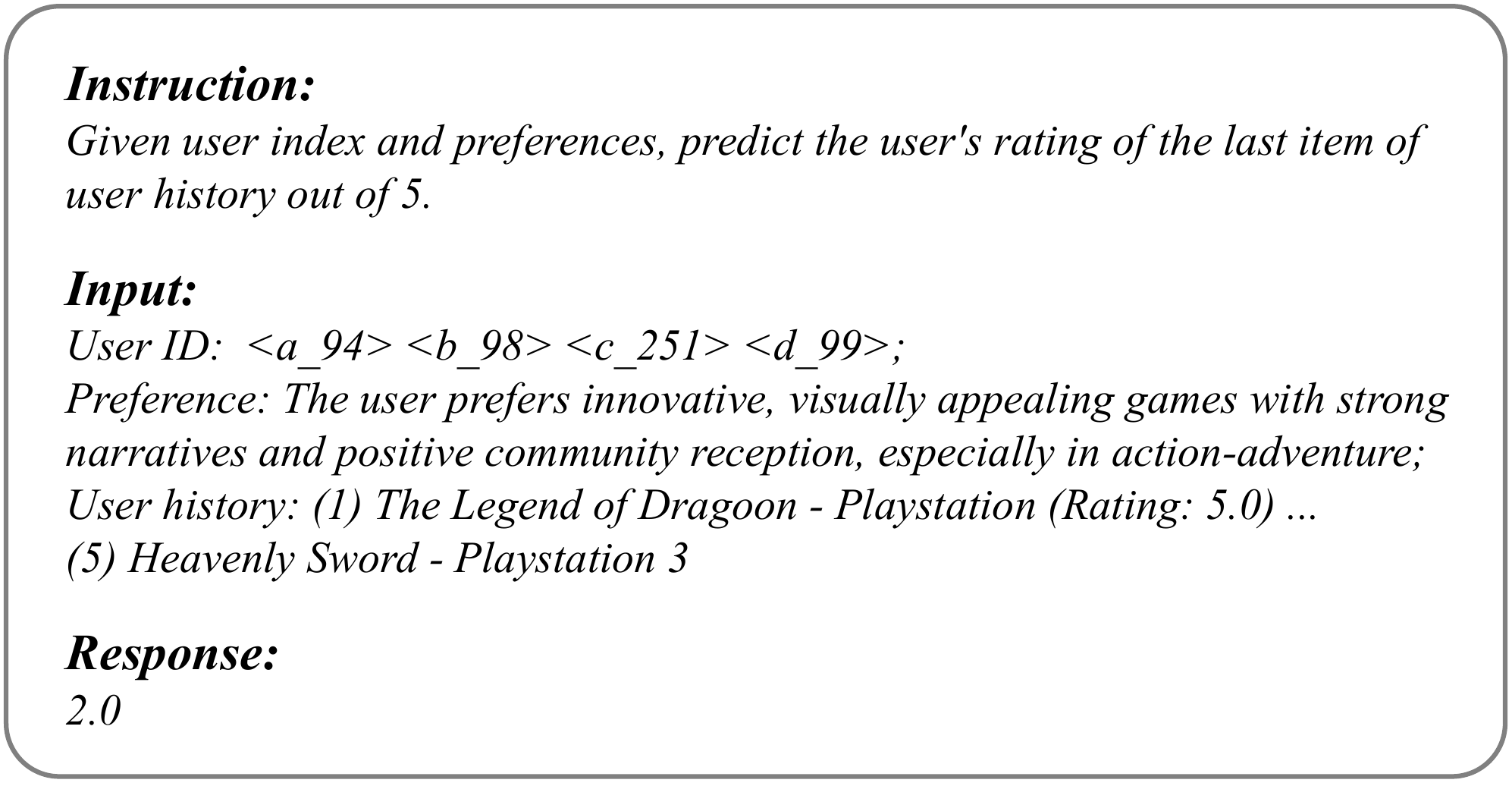}
    \caption{Rating prediction task.}
    \label{ratingprediction}
\end{figure}

To explicitly capture users' preferences towards specific items with greater precision, we incorporate the rating score for deep alignment.
In this task, we provide the LLM with user index, preference, history, and past ratings. 
Notably, the rating for the last interacted item is omitted, leaving it for the LLM to predict. 
This task enhances the LLM's ability to understand and predict fine-grained user preferences, contributing to more personalized recommendations. 
Figure~\ref{ratingprediction} is a sample of the prompt.

\subsubsection{Intent-Based Item Prediction}
\label{sec:intention}
\begin{figure}
    \centering
    \includegraphics[width=0.82\linewidth]{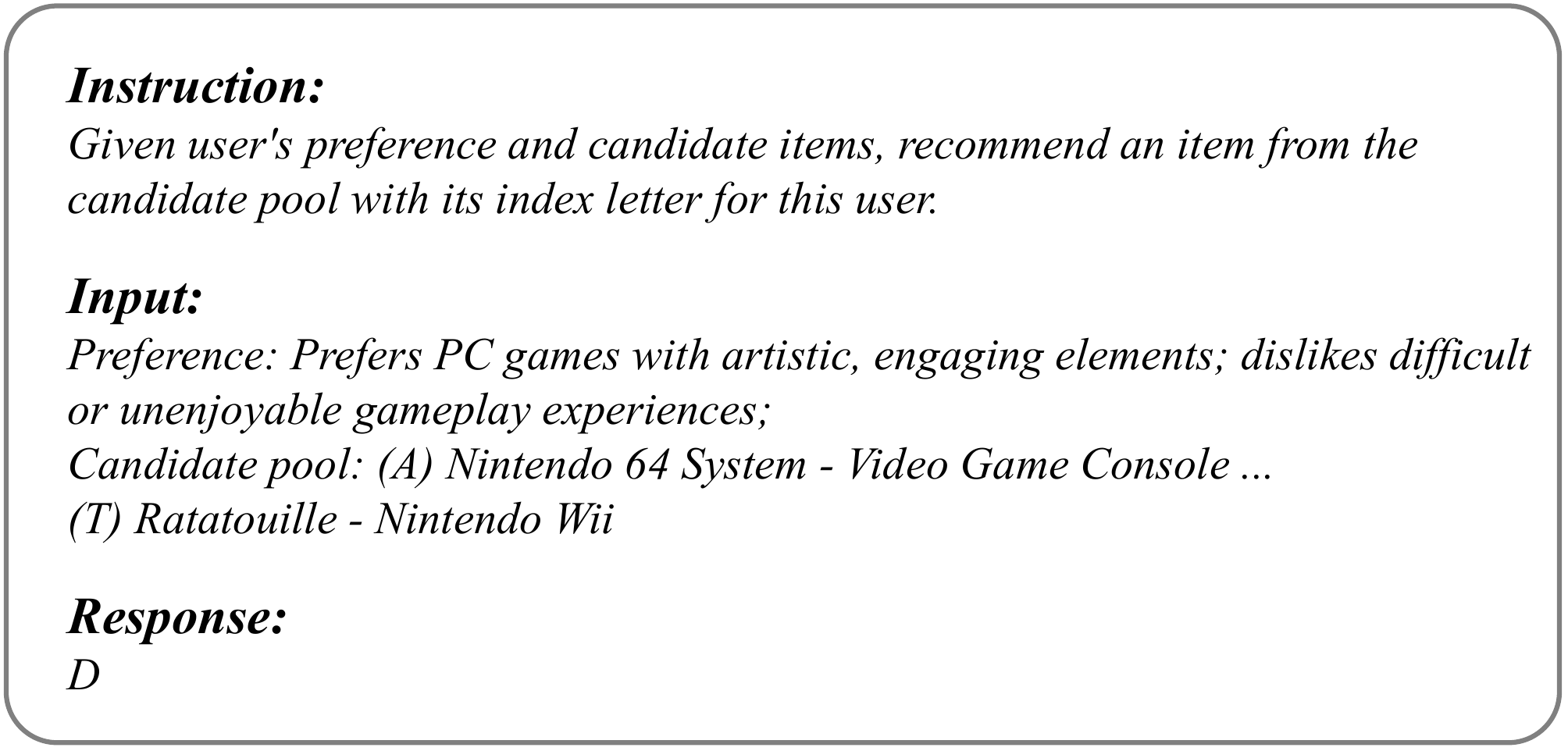}
    \caption{Intent-based item prediction task.}
    \label{intent}
\end{figure}

User preferences generated via GPT-3.5-Turbo contain rich intent information, potentially enabling next-item prediction. 
We hypothesize that the LLM could interpret user intentions from these preferences and make informed recommendations. 
To test it, we design a task in which LLMs are provided with user preferences and candidate items, instructing them to decode user interests and select the most probable preferred item. 
This approach aims to capture nuanced user intentions beyond observable behaviors. 
A sample instance is shown in Figure~\ref{intent}.

\subsection{Model Training and Inference}
During training, our model ranks the candidate items generated by the LRURec~\cite{lrurec} retriever. 
We limit the user history items to a maximum of 20 and select the top-20 candidate items ranked by the retriever model. 
When ranking, we employ Llama-2-7b~\cite{touvron2023llama} as the backbone model and then optimize it through instruction tuning. 
Furthermore, in our implementation, we compute loss solely based on the response section of the prompt.
It is noted that, for task of predicting the next item, we use the verbalizer to transform the LLM head output to a probability distribution without additional parameters. Specifically, it efficiently transforms the output scores over all tokens to ranking scores over candidate items. And we adopt index letters to identify candidate items, mapping the ground truth item to the corresponding index letter like LlamaRec~\cite{llamarec}. Then, the candidate scores can be computed by extracting the logits corresponding to the index letters from the LLM head.

As the objective of all alignments is to generate tokens based on the given context, these tasks are fundamentally identical to the next-token-prediction tasks. 
We optimize the cross-entropy loss of the generation target as follows:
\begin{equation}
{\mathcal L}=-\sum_{i=1}^{m} y_{i} \log \left(\hat{y}_{i}\right) ,
\end{equation}
where $m$ denotes the vocabulary size, $y_{i}$ represents the actual token, and $\hat{y}_{i}$ represents the predicted probability of the token.
In our experiments, we adopt QLoRA~\cite{dettmers2024qlora} to perform quantization on model parameters for efficient training with reduced computation.

During inference, we evaluate the ULMRec performance on the sequential recommendation task with the user index and the user historical behaviors.
Besides, we extracts logits using the verbalizer to rank relevant items in the candidate item pool, which transforms the outputs from the LLM head to ranking scores over the index letters of candidate items for evaluation.
\section{Experiment}

\subsection{Experimental Settings}
\subsubsection{Datasets} We conduct our experiments on two widely used Amazon datasets: \textbf{Beauty} and \textbf{Video Games}~\cite{he2016ups,mcauley2015image}.
Following previous work~\cite{llamarec,lcrec}, we filter out users and items with fewer than five interactions, as well as items lacking titles.
Then, we create user behavior sequences based on the chronological order. 
We report the statistics (i.e., users, items, interactions, sequence length and dataset density) of the processed datasets in Table~\ref{dataset}.

\begin{table}[htbp]
	\centering
	\caption{Statistics of the dataset.}
	\label{dataset}
 \resizebox{0.9\linewidth}{!}{
	\begin{tabular}{cccccc}
	\toprule 
		  Dataset & \# Users &\# Items &\# Interact. &\# Length &\# Sparsity\\ 
		  \midrule 
		  Beauty & 22,332 &12,086 &198k &8.87 &99.9\%\\ 
            Games & 15,264 &7,676 &148k &9.69 &99.8\%\\
	\bottomrule 
	\end{tabular}}
\end{table}

\subsubsection{Baselines}
We compare our method with two different types of baselines:

Classical sequential recommendation models, including preliminary exploration of introducing LLMs into recommendations:
(1) \textbf{NARM}~\cite{narm}: An attention-enhanced RNN model that combines local and global encoders for session-based recommendation.
(2) \textbf{BERT4Rec}~\cite{bert4rec}: A bidirectional Transformer model that learns user preferences through masked item prediction.
(3) \textbf{SASRec}~\cite{sasrec}: A self-attentive sequential model based on unidirectional Transformer architecture.
(4) \textbf{LRURec}~\cite{lrurec}: A lightweight sequential recommender uses linear recurrent units and auto-regressive training.
(5) \textbf{Llama-2}~\cite{touvron2023llama}: We directly use the Llama-2-7b to evaluate the recommendation performance without any fine-tuning.
(6) \textbf{LlamaRec}~\cite{llamarec}: A two-stage framework that uses LRURec for retrieval and fine-tunes Llama-2-7b for list-wise ranking of the pre-selected items.

Recent representative LLM-based sequential recommendation models: (1) Prompt-based methods: a. \textbf{P5}~\cite{p5}: As a text-to-text LLM for recommendation, P5 integrates various tasks and personalized instruction prompts to pre-train the model; b. \textbf{POD}~\cite{pod}: POD converts three recommendation tasks into LLM-based text generation tasks, and then distills continuous prompt vectors from task templates; c. \textbf{PeaPOD}~\cite{peapod}: PeaPOD creates a user-personalized prompt
that distills both the user’s and similar users’ preferences into a learnable soft prompt.
(2) Item-indexing based methods: a. \textbf{TIGER}~\cite{tiger}: A generative retrieval model that utilizes semantic item IDs with Transformer architecture; b. \textbf{CID+IID}~\cite{hua2023index}: The item indexing method based on P5~\cite{p5id}, combining the independent indexing and collaborative indexing; c. \textbf{TransRec}~\cite{transrec}: TransRec utilizes multi-facet identifiers for high-quality identifier generation, and position-free constrained generation to generate ideal items; d. \textbf{IDGenRec}~\cite{idgenrec}: A generative recommender system that incorporates textual ID learning using human language tokens.
(3) User-profile based methods: a. \textbf{PALR}~\cite{palr}: PALR fine-tunes Llama with the user profile keywords generated by user behaviors based on the retrieved candidates; b. \textbf{RDRec}~\cite{rdrec}: Based on POD model, RDRec specifies user and item profiles by distilling interaction rationales from reviews for improving LLM-based recommendation; c. \textbf{P2Rec}~\cite{p2rec}: P2Rec designs a user-level SFT task for collaborative information injection.
Here we present the results with the best-performing backbone model, FMLP-Rec.

\subsubsection{Evaluation}

In evaluation, we employ the leave-one-out strategy: the last item of user history for testing, the second last item for validation, and the rest items for training. 
We adopt three widely used metrics to evaluate different models' performance on sequential recommendation: Mean Reciprocal Rank (MRR@\textit{k}), Normalized Discounted Cumulative Gain (NDCG@\textit{k}) and Recall (Recall@\textit{k}). 
In our experiment, we set \textit{k} as 5 and 10. 
And we use M, N and R to stand for MRR, NDCG and Recall respectively in the following chapters. 
Besides, in the ranking stage, we perform evaluation on the retrieved subset and then combine the ranking metrics with retrieval performance as the overall metrics.

\subsubsection{Implementation Details}

We use the LRURec~\cite{lrurec} for constructing the candidate item pool, which is consistent with the baseline LlamaRec\footnote{https://github.com/Yueeeeeeee/LlamaRec}.
And we generate top-20 candidate items based on the highest Recall@20 metric in the retrieval stage. 
This process is performed with the dropout rate of 0.5 and weight decay of 0.01.
As for the indexing method, we use 4 codebooks to construct user indices, each containing 256 codebook vectors, with each vector having a dimensionality of 32. 
Aligning with LlamaRec, in the ranking stage, we employ Llama2-7b-hf from Huggingface\footnote{https://huggingface.co/meta-llama/Llama-2-7b-hf} as the LLMs backbone for a fair comparison. 
We adopt QLoRA~\cite{dettmers2024qlora} to perform quantization on model parameters with the rank of 8.
We use AdamW optimizer with learning rate as 1e-4 and choose the best model with the highest NDCG@10 on validation set for testing. 
The model is tuned for 1 epoch and validated every 100 iterations.
We conduct each experiment independently, repeating it three times, and report the average results.
All experiments are carried out using the PyTorch framework.
We utilize a single GPU server with an Intel(R) Xeon(R) CPU@2.20GHz and 4 $\times$ NVIDIA A800 GPUs, each with 80GB of memory.

\subsection{Overall Performance}

\subsubsection{Main Performance}

As shown in Table~\ref{mainperformance}, we evaluate the performance of ULMRec and other baseline methods both on valid retrieval subsets and entire datasets.
From the table, we can observe:
(1) Among traditional sequential recommendation models, SASRec consistently outperforms NARM and BERT4Rec across all metrics, underscoring the effectiveness of self-attention mechanisms in capturing sequential dependencies;
(2) The Llama-2 based models show a significant improvement over traditional methods, particularly in the Games dataset, indicating the potential of LLMs in this domain. LlamaRec, which specifically adapts LLMs for recommendation, achieves the best results among baselines across most metrics in both datasets;
(3) Our proposed method demonstrates consistent and substantial improvements over all baselines. For example, in the Games dataset, our method achieves relative improvements of 8.7\%, 8.4\%, and 7.7\% in R@5, R@10, and M@10 respectively, compared to the strongest baseline, LlamaRec. 
These results validate that our method's integration of user personalization into LLMs could effectively capture the user-level preferences beyond item co-occurrence patterns, which could help improve recommendation.

\begin{table*}[htbp]
  \centering
  \caption{Performance of Rank and Overall. Rank refers to the recommendations within the valid retrieval subset, where the ground truth item is among the top 20 retrieved items. Overall refers to the main recommendations within the entire framework. The best and second-best results are in bold and underlined.}
  \resizebox{1.0\linewidth}{!}{
    \begin{tabular}{l|lccc|ccc|ccc|ccc}
    \toprule
    \multicolumn{2}{c}{\multirow{2}[4]{*}{\textbf{Method}}} & \multicolumn{6}{c}{\textbf{Rank}}             & \multicolumn{6}{c}{\textbf{Overall}} \\
\cmidrule(r){3-8} \cmidrule(r){9-14}    \multicolumn{2}{c}{} & \textbf{M@5} & \textbf{N@5} & \multicolumn{1}{c}{\textbf{R@5}} & \textbf{M@10} & \textbf{N@10} & \multicolumn{1}{c}{\textbf{R@10}} & \textbf{M@5} & \textbf{N@5} & \multicolumn{1}{c}{\textbf{R@5}} & \textbf{M@10} & \textbf{N@10} & \textbf{R@10} \\
    \midrule
    \multirow{6}[2]{*}{Games} & NARM  & 0.2039  & 0.2424  & 0.3600  & 0.2248  & 0.2931  & 0.5168  & 0.0479  & 0.0576  & 0.0874  & 0.0541  & 0.0729  & 0.1351  \\
          & BERT4Rec  & 0.1765  & 0.2109  & 0.3160  & 0.1947  & 0.2551  & 0.4530  & 0.0422  & 0.0512  & 0.0788  & 0.0478  & 0.0649  & 0.1214  \\
          & SASRec   & 0.2177  & 0.2571  & 0.3776  & 0.2408  & 0.3134  & 0.5521  & 0.0515  & 0.0617  & 0.0930  & 0.0583  & 0.0783  & 0.1446  \\
          &  Llama-2 & 0.2264  & 0.2720  & 0.4117  & 0.2558  & 0.3439  & 0.6352  & 0.0477  & 0.0574  & 0.0868  & 0.0539  & 0.0725  & 0.1339  \\
          & LRURec   & 0.2504  & 0.3009  & 0.4544  & 0.2811  & 0.3760  & 0.6879  & 0.0533  & 0.0640  & 0.0966  & 0.0598  & 0.0800  & 0.1463  \\
          & LlamaRec & \uline{0.2825}  & \uline{0.3360}  & \uline{0.4995}  & \uline{0.3158}  & \uline{0.4173}  & \uline{0.7522}  & \uline{0.0600}  & \uline{0.0714}  & \uline{0.1061}  & \uline{0.0671}  & \uline{0.0887}  & \uline{0.1599}  \\
          & Our   & \textbf{0.3071 } & \textbf{0.3641 } & \textbf{0.5379 } & \textbf{0.3364 } & \textbf{0.4354 } & \textbf{0.7592 } & \textbf{0.0647 } & \textbf{0.0768 } & \textbf{0.1134 } & \textbf{0.0709 } & \textbf{0.0918 } & \textbf{0.1600 } \\
    \midrule
    \multirow{6}[2]{*}{Beauty} & NARM  & 0.1961  & 0.2284  & 0.3263  & 0.2128  & 0.2689  & 0.4517  & 0.0289  & 0.0342  & 0.0503  & 0.0321  & 0.0420  & 0.0746  \\
          & BERT4Rec  & 0.1587  & 0.1901  & 0.2861  & 0.1743  & 0.2281  & 0.4043  & 0.0246  & 0.0298  & 0.0457  & 0.0276  & 0.0372  & 0.0686  \\
          & SASRec   & 0.2296  & 0.2679  & 0.3843  & 0.2491  & 0.3152  & 0.5312  & 0.0336  & 0.0397  & 0.0582  & 0.0371  & 0.0481  & 0.0844  \\
          & Llama-2 & 0.2617  & 0.3047  & 0.4360 & 0.2876  & 0.3687 & 0.6365 & 0.0344 & 0.0401 & 0.0574 & 0.0378  & 0.0485  & 0.0837 \\
          & LRURec   & 0.2944  & 0.3403  & 0.4801  & 0.3259  & 0.4168  & 0.7170  & 0.0376  & 0.0435  & 0.0614  & 0.0417  & 0.0533  & 0.0916  \\
          & LlamaRec & \uline{0.3016}  & \uline{0.3524}  & \uline{0.5071}  & \uline{0.3350}  & \uline{0.4337}  & \textbf{0.7600 } & \uline{0.0385}  & \uline{0.0450}  & \uline{0.0648}  & \uline{0.0428}  & \uline{0.0554}  & \uline{0.0971 } \\
          & Our   & \textbf{0.3253 } & \textbf{0.3792 } & \textbf{0.5436 } & \textbf{0.3520 } & \textbf{0.4445 } & \uline{0.7468}  & \textbf{0.0428 } & \textbf{0.0499 } & \textbf{0.0715 } & \textbf{0.0463 } & \textbf{0.0585 } & \textbf{0.0982 } \\
    \bottomrule
    \end{tabular}}%
  \label{mainperformance}%
\end{table*}%

\subsubsection{Further Comparison with LLM-based Recommenders}
Furthermore, we compare the performance of our model with recent representative LLM-based methods in Table~\ref{overallperformance}. 
Similar to LlamaRec, we compare against the reported results from their original papers, as the data processing and evaluation methods are nearly identical. 
From the results, we can find: 
(1) P5, POD and PeaPOD propose innovative solutions that yield commendable results, but the extent of these improvements remains limited. The representation of original IDs presents challenges for LLMs in interpreting the naturally discrete ID indexing. And the inherent collaborative semantics and user preference information may be potentially destroyed during tokenization;
(2) TIGER, CID+IID, TransRec and IDGenRec utilize item indexing methods to enhance sequential recommendation performance. However, limited by merely focusing on the item-related information, they neglect the high-order user personalized preference, which is also crucial to recommendation;
(3) PALR and RDRec introduce user profile into recommendation for realizing competitive results. 
PALR injects the summarized preference from user history into the tuning prompt.
RDRec leverages user preference and item attribute generation task to train the LLM.
However, these methods would result in users not having a unique identifier to refer to them, which may cause confusion about different user personalized preferences;
(4) Our ULMRec shows significant relative improvements compared to the second best results, achieving 32\%, 19\%, 16\%, and 8\% improvements respectively in R@10, R@5, N@10, and N@5, which validates the effectiveness of the ULMRec.

\begin{table*}[htbp]
  \centering
  \caption{The overall performance compared to other LLM-based models on Beauty dataset.}
  \resizebox{0.84\linewidth}{!}{
    \begin{tabular}{cccccccccccc}
    \toprule
          & P5    & PALR  & TIGER  & CID+IID & TransRec & POD  & PeaPOD  & RDRec  & P2Rec  & IDGenRec  & Our \\
    \midrule
    \textbf{N@5} & 0.0367 & N/A   & 0.0321   & 0.0356 & 0.0365 & 0.0395 & 0.0445 & 0.0461 & 0.0445 & \uline{0.0486} & \textbf{0.0499 } \\
    \textbf{R@5} & 0.0493 & N/A & 0.0454   & 0.0512 & 0.0504 & 0.0537 & 0.0588 & 0.0601 & 0.0604 & \uline{0.0618} & \textbf{0.0715 } \\
    \midrule
    \textbf{N@10} & 0.0416 & 0.0446 & 0.0384  & 0.0427 & 0.0450 & 0.0443 & 0.0493 & 0.0504 & 0.0509 & \uline{0.0541} & \textbf{0.0585 } \\
    \textbf{R@10} & 0.0645 & 0.0721 & 0.0648 & 0.0732 & 0.0735 & 0.0688 & 0.0738 & 0.0743 & \uline{0.0852} & 0.0814 & \textbf{0.0982 } \\
    \bottomrule
    \end{tabular}}%
  \label{overallperformance}%
\end{table*}%

\subsection{Ablation Study}
\label{sec:ablation}

To further explore how different alignment tasks impact the model performance, we design six variants for experimentation (Games dataset as examples), including:
(1) w/o pref: removing the index-preference alignment task in ~\ref{sec:pref},
(2) w/o intention: removing the intent-based item prediction task in ~\ref{sec:intention},
(3) w/o rating: removing the rating prediction task in~\ref{sec:rating},
(4) w/o history: removing the history-index alignment task in~\ref{sec:history},
(5) w/o turnpref: removing the preference-index alignment task in~\ref{sec:turnpref}.
(6) w/o all-align: removing all alignment tasks except the next-item prediction task.

\begin{figure}
    \centering
    \includegraphics[width=0.77\linewidth]{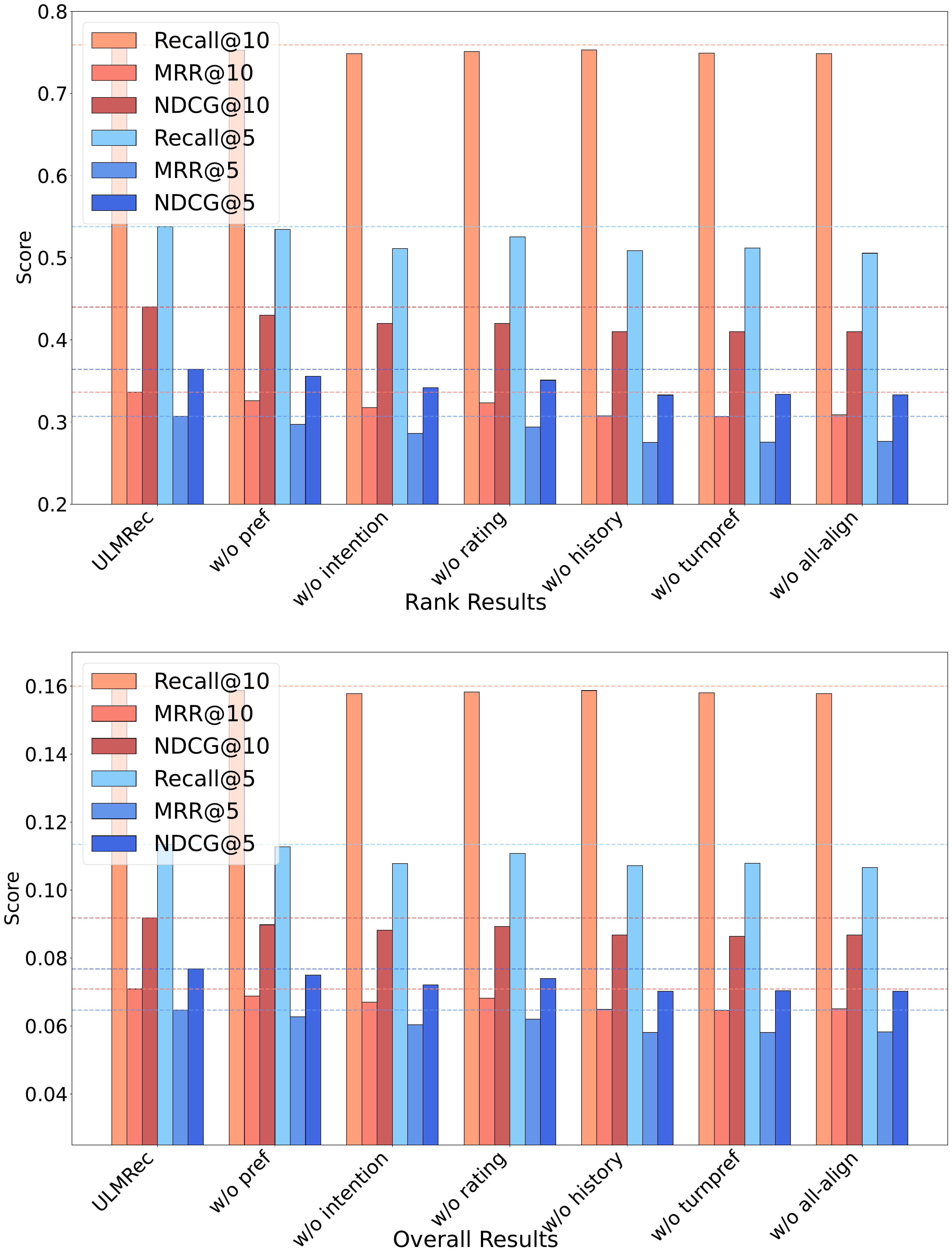}
    \caption{Ablation study on the alignment tasks.}
    \label{ablationphoto}
\end{figure}

The experimental results are shown in Figure~\ref{ablationphoto}, we can find each alignment task contributes to recommendation performance improvements, demonstrating the effectiveness of our designed preference alignment tasks.
It is not surprising that w/o all-align obtains the poorest performance, this is because that LLMs could not effectively capture user personalized preferences solely based on the next-item prediction task.
Besides, we find the performance of w/o turnpref is worse than that of w/o pref.
This may be because that w/o turnpref taking the user's personalized preferences as input makes it easier for the LLMs to understand and generate the corresponding user index.
In all, all these instruction tuning tasks are shown beneficial for enhancing LLM-based sequential recommendation.


\subsection{Further Analysis}

\subsubsection{The user intentions in index}
To evaluate our indexing method's ability to capture user intentions, we conduct experiments using our trained model to predict the next item based on the user index and a candidate item pool, without chronological user history (named ULMRec-uid). 
This ensures the LLM relies entirely on learned user semantic preferences through indexing.
As shown in Table~\ref{tab:intention} and Table~\ref{mainperformance}, the well-trained LLM using the aligned index outperforms traditional models like BERT4Rec, especially in the R@10 metric. 
This superior performance, achieved without historical interaction data, demonstrates that ULMRec could effectively capture personalized user interests based on the user index alone.
These results validate our indexing method's effectiveness in creating comprehensive user profiles and highlight the potential of LLM-based approaches in understanding user preferences, even in scenarios with limited historical data.

\begin{table}
\centering
\caption{Recommendation performance based solely on user index and candidates, excluding chronological history (Games dataset in ranking phase as the example).}
\resizebox{0.49\textwidth}{!}{
\begin{tabular}{c|c|c|c|c|c|c}
\toprule[1pt]
\diagbox{Model}{Metric} & M@5 & N@5 & R@5 & M@10 & N@10 & R@10 \\
\midrule
ULMRec & 0.3071 & 0.3641 & 0.5379 & 0.3364 & 0.4354 & 0.7592 \\
\midrule
ULMRec-uid & 0.1843 & 0.2330 & 0.3822 & 0.2209 & 0.3233 & 0.6647 \\
\bottomrule[1pt]
\end{tabular}
}
\label{tab:intention}
\end{table}

\begin{figure}
    \centering    \includegraphics[width=0.85\linewidth]{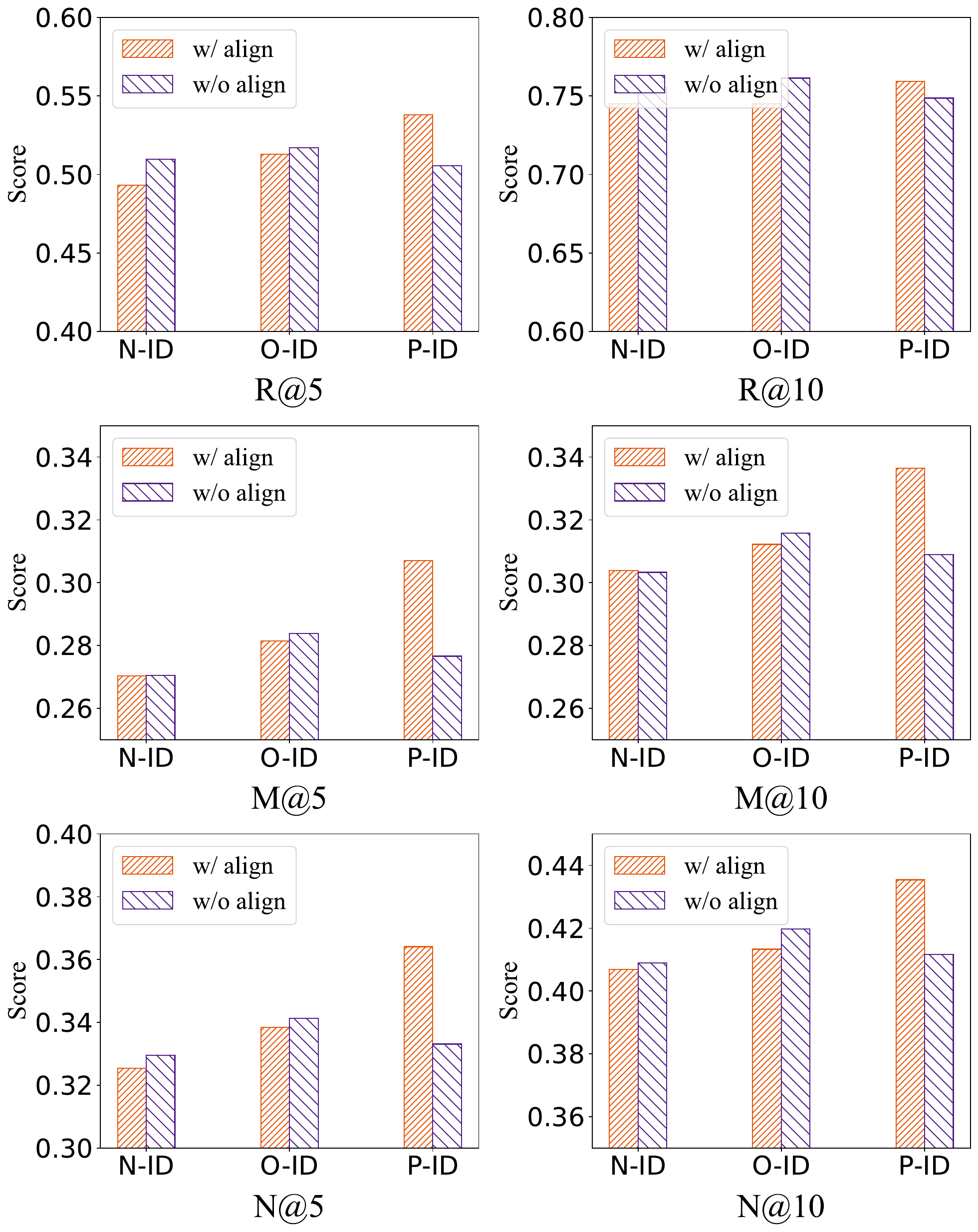}
    \caption{Performance comparison on different user indexing methods: 1) N-ID: Numerical IDs. 2) O-ID: Original IDs. 3) P-ID: Our personalized IDs.}
    \label{id}
\end{figure}

\subsubsection{Different User Indexing Methods}
\label{sec:id}
To comprehensively evaluate the effectiveness of our proposed user indexing method, we conduct a comparative analysis of three distinct user indexing approaches.
(1) Numerical IDs (N-ID): conventional recommendation user numerical identifiers (e.g., 1, 2, 3); (2) Original IDs (O-ID): unmodified user identifiers from the dataset (e.g., A1GNYV0RA0EQSS), typically alphanumeric strings; (3) Our personalized user indexing method (P-ID): personalized vector quantization approach based on user reviews and IDs (e.g., <a\underline{~}219> <b\underline{~}2> <c\underline{~}95> <d\underline{~}238>).
Additionally, we investigate the impact of alignment tasks on each indexing method to assess their interaction with the LLM (i.e., w/ align and w/o align).
As illustrated in Figure~\ref{id}, our VQ-based method with alignment consistently outperforms other approaches across all metrics, demonstrating the effectiveness of our designed user indexing method.
It's worth mentioning that for N-ID or O-ID, models without alignment perform better. 
This may be because that the N-ID and O-ID would be understood by the LLMs as the original text semantics and fail to effectively represent user personalized preference.
Besides, the additional alignment tasks would also destroy the original semantics and lead to performance degradation.
In contrast, our personalized VQ-based user indexing method and carefully designed alignment tasks could enable the model to effectively integrate the user personalized preference into the LLMs.

\subsubsection{Complexity}
In this section, we analyze the complexity of ULMRec. 
We employ GPT-3.5-Turbo to generate and store user preferences, a one-time operation per user. 
Subsequently, we fine-tune the model based on the Llama-2 architecture, requiring a single Llama-2 call per data point.
For inference, ULMRec necessitates only one Llama-2 call per data instance.
Notably, compared to other LLM-based recommendation models, ULMRec maintains a similar number of LLM requests, demonstrating comparable efficiency.
On the Beauty dataset, ULMRec averages 0.615 seconds per test data point, slightly faster than LlamaRec's 0.636 seconds. 
For the Games dataset, ULMRec takes 0.423 seconds, marginally slower than LlamaRec's 0.417 seconds. 
These results indicate that ULMRec maintains comparable inference times to other LLM-based models.

\subsubsection{Case Study}
\begin{figure}
    \centering
    \includegraphics[width=0.86\linewidth]{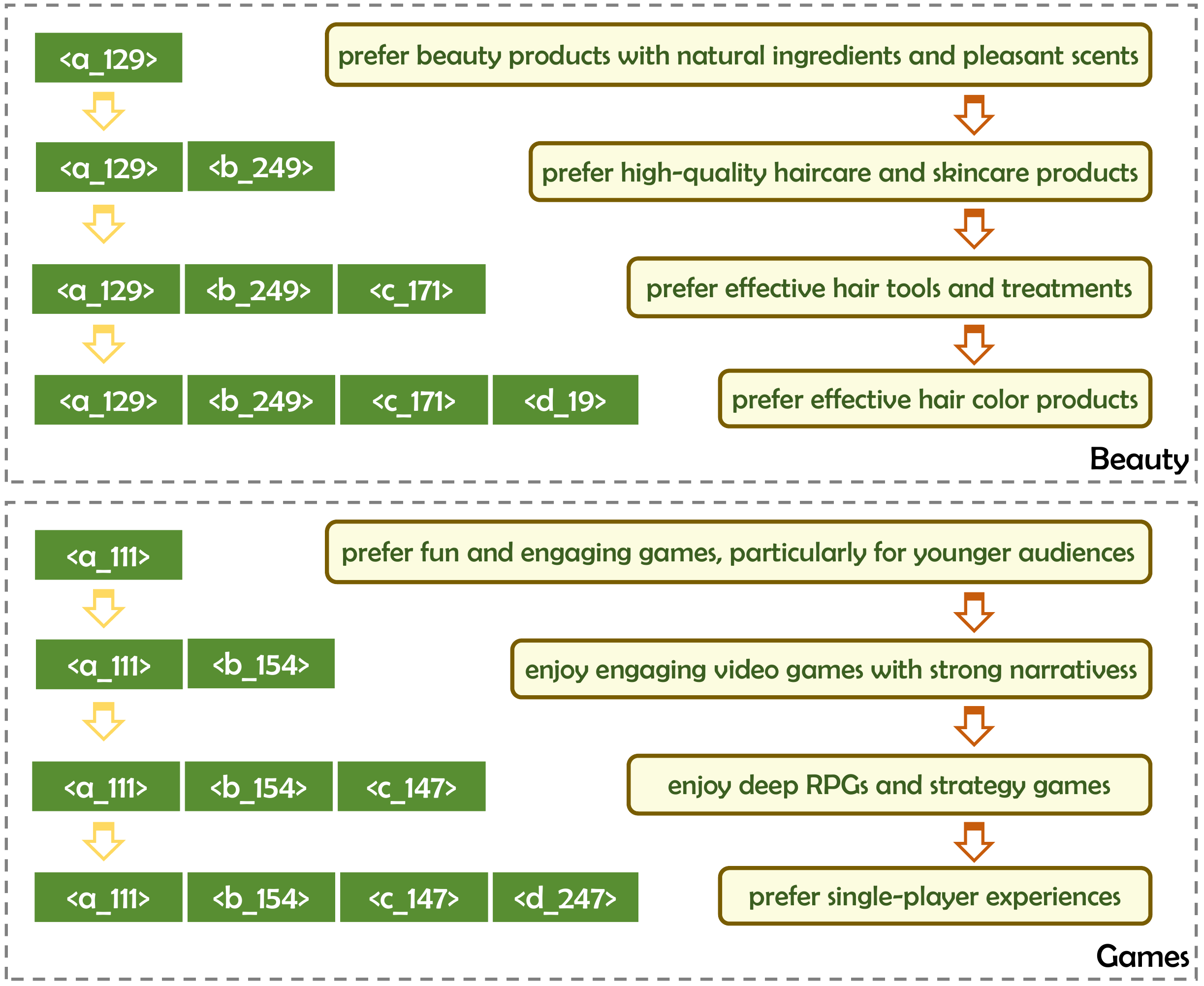}
    \caption{Case study for the preference with user index.}
    \label{case}
\end{figure}

To further explore the relationship between user preference and indices, we present two illustrative cases in Figure~\ref{case}.
We prompt the trained model with different hierarchical indices and enforce it to predict the user preference. As the level of indices increases, we observe that the predicted preference features become more specific, evolving from broad categories to precise item types. For instance, in the Beauty dataset, the first-level index only refers to a general preference, the second-level narrows it down to haircare products, and the third-level further focuses on the hair tools. When the user information is complete, the model can predict the specific interests towards hair color products. 
This progression demonstrates the effectiveness of the alignment between preference information and user index.

\section{Conclusion}

We introduce ULMRec, an LLM-based recommender for sequential recommendation that integrates user-item interactions and user personalized information. 
Our approach generates unique semantic user indices through vector quantization of review embeddings and user IDs, then employs alignment tasks to incorporate user-specific semantics. These tasks include sequential recommendation, explicit and implicit alignments, enabling LLMs to map indices to user characteristics and bridge semantic gaps. Experiments demonstrate ULMRec's effectiveness in both indexing and alignment, outperforming state-of-the-art models in recommendation. 
In future work, we will explore the transferability of our LLM-learned user indices across diverse recommendation tasks and domains, potentially opening new avenues for cross-domain personalization.

\clearpage

\bibliographystyle{ACM-Reference-Format}
\normalem
\bibliography{www}

\end{document}